\documentclass[12pt]{article}
\usepackage{amsfonts,amsmath}
 \usepackage{amssymb}
\usepackage[hypertex] {hyperref}

\makeatletter \@addtoreset{equation}{section} \makeatother

\newcommand{\tgo} { \Lambda_N(\eta)}
\newcommand{\gtgo} { \Lambda }
\newcommand{\hg}{ {{g}}}
\newcommand{\vf} {  {f}}

\newcommand{\hgg}{ {\rm{g}}}
\newcommand{\vff} {  {\rm{f}}}

\newcommand{\dis}{\displaystyle}

\newcommand{\dr}{{\rm d}}
\newcommand{\nv}{\,n}
\newcommand{\kv}{\,k}
\newcommand{\mv}{\,m}
\newcommand{\mh}{\,\widehat{m}}
\newcommand{\nh}{\,\widehat{n}}
\newcommand{\kh}{\,\widehat{k}}
\newcommand{\ph}{\,\widehat{p}}

\newcommand{\plh}{ \widehat{+}}
\newcommand{\mih}{ \widehat{-}}

\newcommand{\gdp}{\kappa}
\newcommand{\bgdp}{\zeta}

 \newcommand{\ua}{{\underline{A}}}
\newcommand{\ub}{{\underline{B}}}

\newcommand{\la}{ {{a }}}
\newcommand{\lb}{ {{b} }}
\newcommand{\lap}{ {{a }'}}

\newcommand{\lbp}{ {{b}' }}

 \newcommand{\ad}{\tau}
 \newcommand{\bd}{\upsilon}

\newcommand{\adm}{\vartheta}
 \newcommand{\bdm}{\varphi}

\newcommand{\co}{{\rm{  c }}}
\newcommand{\si}{{\rm{  s }}}

\newcommand{\BB}{{\cal B}}
\newcommand{\BBB}{{\mathfrak{B}}}

\newcommand{\be}{ \begin{equation}}

\newcommand{\etc}{{\it etc}}
\newcommand{\slv}{{}^v{}\mathfrak{su}(2)}
\newcommand{\slh}{{}^h{}\mathfrak{su}(2)}

\newcommand{\bp}{\bar{\partial}}

\newcommand{\ee}{\end{equation}}
 \newcommand{\PPP}{ {J} }

\newcommand{\bee}{\begin{eqnarray}}
\newcommand{\beee}{\begin{array}}
\newcommand{\eee}{\end{eqnarray}}
\newcommand{\eeee}{\end{array}}
\newcommand{\hhh}{{\hbar}}

 \newcommand{\gm}{\mu}
\newcommand{\gnn}{\nu}
 \newcommand{\gmm}{\mu}

\newcommand{\gx}{\xi}

\newcommand{\ga}{\alpha}
 \newcommand{\pa}{{\ga^\prime}}
 
\newcommand{\pn}{{\nu^\prime}}

\newcommand{\pb}{{\gb^\prime}}
\newcommand{\pga}{{\gamma^\prime}}
\newcommand{\gb}{\beta}

\newcommand{\gga}{\gamma}

\newcommand{\pgga}{{\gamma^\prime}}

\newcommand{\M}{{\cal M}}

\newcommand{\G}{{\cal G}}

\newcommand{\Hh}{{\cal H}}

\newcommand{\ie}{{\it i.e.,} }
\newcommand{\ls}{\!\!\!\!\!\!}

\newcommand{\gd}{\delta}

\newcommand{\gl}{\lambda}

\newcommand{\gvep}{\varepsilon}

\newcommand{\bz}{\bar z}
\newcommand{\go}{\omega}
\newcommand{\goL}{\omega^{ L}}

\newcommand{\bgoL}{\bar{\omega}^{  L}}

\newcommand{\by}{{\bar{y}}}

\newcommand{\q}{\,,\qquad}

\newcommand{\nn}{\nonumber}

\newcommand{\chalf}{\frac{1}{4}}
\newcommand{\half}{\frac{1}{2}}

\newcommand{\p}{\partial}

\newcommand{\f}{\frac}
\newcommand{\A}{{\cal A}}

\newsavebox{\ver}
\newsavebox{\verp}
\newsavebox{\gorp}
\newsavebox{\toch}

\begin{document}

\begin{flushright}
\vspace{1mm}
 FIAN/TD/2014-22\\
 December {2014}\\
\end{flushright}

\vskip1.5cm

 \begin{center}
 {\large\bf
  Conserved higher-spin charges in $AdS_4$}
 \vglue 0.6  true cm

\vskip0.5cm

 O.A. Gelfond$^1$ and M.A.~Vasiliev$^2$
 \vglue 0.3  true cm

 ${}^1$Institute of System Research of Russian Academy of Sciences,\\
 Nakhimovsky prospect 36-1,
 117218, Moscow, Russia

 \vglue 0.3  true cm

 ${}^2$I.E.Tamm Department of Theoretical Physics, Lebedev Physical
 Institute,\\
 Leninsky prospect 53, 119991, Moscow, Russia

 \end{center}

\vskip2cm

 \begin{abstract}
 Gauge invariant conserved conformal currents built from massless fields of
 all spins  in 4d Minkowski space-time and $AdS_4$  are described in the
 unfolded dynamics approach. The current cohomology associated with  non-zero
 conserved charges is found.  The resulting list of charges is shown to match the
 space of parameters of the conformal higher-spin symmetry algebra in four
 dimensions.

   \end{abstract}
\newpage

\tableofcontents

\section{Introduction}

Gauge invariant conserved currents of different spins in $4d$ Minkowski space
were constructed in \cite{Berends:1985xx}  in
terms of generalized higher-spin (HS) Weyl curvatures introduced originally
in \cite{Weinberg:1965rz}. The latter describe on-shell nontrivial gauge invariant
combinations of derivatives of fields which generalize the spin-one Maxwell
tensor and linearized spin-two Weyl tensor.  Later conserved HS currents were
also considered
  in \cite{Anco:0202019,GSV} while nontrivial
currents leading to non-zero charges were identified in \cite{LS}.

 In \cite{BHS} it was shown that global conformal HS  symmetries of $4d$ massless fields of all spins are described by the Weyl algebra $A_4$
 of eight oscillators.  Algebras of symmetries of equations of motion
 of irreducible free fields and supermultiplets were also found in \cite{BHS}
 while extensions to higher dimensions were elaborated in
 \cite{Eastwood:2002su,Vasiliev:2004cm,Bekaert:2009fg,Govil:2014uwa}.

Closed
forms describing the gauge invariant conservation laws in $4d$  Minkowski
space-time    were found within the unfolded approach in \cite{GSV}. In this paper
we extend these results  to  $AdS_4$ and analyze the current
cohomology characterizing nontrivial conserved charges. Namely, in \cite{GVC}
it was shown that the space of closed three-forms which can give rise to
conserved charges is far larger than the space of HS conserved
charges that can be associated with symmetries of massless fields. Hence,
it was conjectured in \cite{GVC} that most of such closed three-forms
are exact. In this paper, we show that this is indeed true and that the
current cohomology
 matches the anticipated HS global symmetries.
We focus on the gauge invariant currents built in terms of generalized Weyl
tensors. Note that non-gauge invariant conserved currents
 built in terms of HS connections are also
available \cite{Smirnov:2013kba} in $4d$ Minkowski space and $AdS_4$ \cite{Smirnov:2015}
where it was also shown that the conserved charges associated with these currents are
gauge invariant.

A convenient way to analyze  conserved quantities is provided by  the
so-called unfolded approach
\cite{4dun}  in which all fields and field equations are formulated in terms
of differential forms (for a review see e.g. \cite{Bekaert:2005vh}).
 The full set of nontrivial symmetry parameters of bilinear conserved
 currents, that lead to non-zero charges,  was found
in  \cite{BRST2010} using  the correspondence between
unfolded and BRST formulations of $Sp(2M)$ invariant HS
equations in the generalized matrix space-time $\M_M$.
For $M=2$ the respective charges were shown in \cite{OPE}  to   generate the
$3d$ $N=2$ superconformal HS algebra equivalent to the $N=2$ $AdS_4$ HS algebra.

An important advantage of the unfolded formulation is that in lower dimensions
$d=3,4$  dynamical content of a theory is fully characterized by auxiliary spinor
variables $Y$.  Transition from fields $C(Y|x)$ to currents $J(Y_1,Y_2|x)$
corresponds to the tensor product of the  modules associated with the fields $C(Y|x)$.
Technically, all properties of a system are encoded by the $Y$-dependence that
determines in particular the space-time evolution of the system in question. As a result,
 properties of modules in the $Y$-space
characterize properties of the related physical quantities in the $x$-space. In the context
of the present paper this gives us an opportunity to distinguish between nontrivial and
trivial conserved currents in terms of certain algebras acting in the $Y$-space.
Specifically, we identify a Lie algebra   $\mathfrak{o}(4,4)$
 that   maps any
closed three-form associated with a bilinear conserved
current in $AdS_4$ to  an exact form, \ie improvement. As a result,
factorization of the generators of this $\mathfrak{o}(4,4)$ allows one to factor
out exact conserved currents. The resulting non-exact  bilinear conserved currents in $AdS_4$ are given in   (\ref{concur2helCC}).
Analogous results for Minkowski space
 are obtained in Section \ref{Currents Min} (see
  (\ref{concur2hel})).

Extension of the construction of gauge invariant conserved charges to $AdS_4$
performed in this paper may have various applications including in
particular  computation of HS charges of black hole-like solutions in HS theory
\cite{blackH,Iazeolla:2011cb}. The latter are stationary spherically symmetric solutions
of full nonlinear HS field equations that have a singularity in the bulk and fall down
properly at space infinity of $AdS_4$ space-time. Such solutions are analogous to usual
GR black holes  although their fundamental properties
still remain largely unexplored (it is not even
clear what kind of horizon (if any) is associated with them). The conserved
HS currents introduced in this paper can be used for the perturbative
analysis of HS black hole charges. Being associated with closed three-forms,
the related charges  result from integration over the three-dimensional
bulk space. This is different from the construction of two-form currents giving rise to
asymptotic charges in $AdS$ as proposed e.g. in \cite{AMD1,AMD2}. (Two-form
charges in HS theory were discussed recently in \cite{vasapr,DMV}). On the other hand the
both constructions are related because in the area of their applicability the
two types of charges coincide pretty much the same way as in the Gauss
law in electrodynamics. More generally,
the fields contributing to the asymptotic charges
are sourced by the currents constructed in this paper, although,
being simple at the linearized level, the relation between the two types of charges
is far less trivial in the full nonlinear theory.

The rest of the paper is organized as follows. In Section \ref{Higher-rank fields} the description of higher-rank fields in the unfolded approach is
briefly recalled. In Section \ref{Rank-two equations} we recall the unfolded
form of free 
 HS  equations in $AdS_4$  proposed in
\cite{4dun,Ann} and their flat limit. In Section \ref{Currents AdS} conserved
HS currents in $AdS_4$  are constructed in terms of
covariantly-constant oscillators and De Rham cohomology of gauge-invariant
conserved conformal currents built from massless fields of all spins is
found.  The resulting nontrivial charges are shown to
 match the space of parameters of the HS symmetry algebra.
In Section \ref{Currents Min} conserved currents of $4d$ Minkowski space are
reconstructed via the flat limit of those in $AdS_4$.

 \section{Higher-rank fields}
 \label{Higher-rank fields}
Conformal massless fields of all spins in four dimensions can be described
\cite{BHS} by a rank--one zero-form $C(Y|x)$ where $x^n$ are $4d$ space-time coordinates and    $Y^A$ are auxiliary spinor
variables ($A,B=1,...,4\,$ are Majorana spinor indices). It is convenient to
interpret $C(Y|x)$ as a vector $ |C(Y|x)\rangle $ in the Fock space $F$
of the algebra   of oscillators $Y^A$ and $Z_A$ that satisfy commutation
relations \be \label{cryz} [Y^A\,,Y^B]=0\q [Z_A\,,Z_B]=0\q [Z_A\,,Y^B]
=\delta_A^B\,. \ee The Fock vacua $|0\rangle$ and $\langle0| $ are defined to
obey \be\label{rangle} Z_A |0\rangle =0\,\q
 \langle0| Y^A   =0\,. \ee
In these terms the rank--one equation of \cite{BHS} takes the form \be
\label{frank1} D|C(Y|x)\rangle := \big(\dr+ W(Y,Z|x)\big)|C(Y|x)\rangle =0\,,
\ee where $\dr = dx^n \f{\p}{\p x^n}$ is De Rham derivative and
$W(Y,Z|x)$ satisfies the flatness condition \be \label{dW} D^2
=0\,:\qquad \dr W +\half [W \,,\wedge  W ]=0\,. \ee

  It is convenient to choose  $W(Y,Z|x)$ valued in the
  $\mathfrak{sp}(8)$ realized by bilinears of $Y^A$ and $Z_A$
\bee\label{Wrank1} W(Y,Z|x)=     f_{A}{}_{B}(x)\, Y^A Y^B +  h^{AB}(x)\, Z_A
Z_B + \half\go_{A}{}^{B}(x)\, \{ Y^A \,,\,Z_B\} \q \eee where $ h^{AB}$, $
f_{A}{}_{B}$ and $ \go_{A}{}^{B}$ are components of the one-form connection.
The
 $\mathfrak{sp}(8)$ flatness conditions are
 \bee
\label{Rsp8} &&R^{AB}:= \dr\,h^{AB} -\omega_C{}^A\wedge h^{CB}  =0\q R_{AB}:=
\dr\,f_{AB} +\omega_A{}^C\wedge f_{CB}  =0\,,
\\ \nn
&&R_A{}^B := \dr\,\omega_A{}^B +\omega_A{}^C\wedge \omega_C{}^B -
 f_{AC}\wedge h^{CB}=0\,. \eee

From the oscillator realization it is obvious that   massless field equations
(\ref{frank1}) are invariant under the global symmetry associated with the
full Weyl algebra of the oscillators $Y$ and $Z$. Indeed, suppose that
$|C(Y|x)\rangle$ solves  rank--one equations (\ref{frank1}). Then any
$|\widetilde{C}(Y|x)\rangle$ of the form \be |\widetilde
C(Y|x)\rangle=\eta(Y,Z|x) |C(Y|x)\rangle \ee with $\eta(Y,Z|x)$ satisfying
\be \label{r1sym} D\eta(Y,Z|x)  := \dr \eta(Y,Z|x) +[W(Y,Z|x),\eta(Y,Z|x)] =0
\ee also solves (\ref{frank1}). Since  equations (\ref{r1sym}) are
consistent with $D^2=0$ by virtue of  flatness condition (\ref{dW}), their
general solution is reconstructed uniquely in terms of $\eta(Y,Z|x_0)$ at any
given point $x_0$ (denoted by 0 in the sequel),
hence being characterized by an arbitrary function of $Y$\! and $Z$.

Because $W(Y,Z|x)$ is bilinear in the oscillators $Y$ and $Z$,
equation (\ref{r1sym}) is homogeneous in the oscillators. In particular, one
can solve Eq.~(\ref{r1sym}) for $\eta(Y,Z|x)$ linear in $Y$ and $Z$. Clearly,
there are eight independent solutions of this type which we denote

\be\label{AaBb}\A^\ua (Y,Z|x)\q\BB_\ub (Y,Z|x)\,,\quad\ee
 where $\ua, \ub =1,\ldots 4$ label independent solutions
normalized so that
\be\label{normA} \A^A (Y,Z|0) = Y^A\q \BB_A (Y,Z|0) = Z_A\,.
\ee
 This normalization guarantees
 that $\A^\ua(x)$ and $\BB_\ub(x)$ obey canonical commutation
relations  at any $x$ \be \label{crab} [\A^\ua\,,\A^\ub]=0\q
[\BB_\ua\,,\BB_\ub]=0\q [\BB_\ua\,,\A^\ub] =\delta_\ua^\ub\,. \ee Indeed, since
the oscillators $\A^\ua$ and $\BB_\ub$ are covariantly constant with respect
to $D$, their commutator is also covariantly constant. Since the commutator
of linear combinations of the oscillators $Y^A$ and $Z_B$ is independent of
the oscillators, $D$ acts on the commutator as  $\dr$, and hence the right-hand-sides of (\ref{crab})
should be coordinate independent.

In terms of $\A^\ua$ and $\BB_\ua$, general solution of  equation
(\ref{r1sym}) is
\be\label{sol-osc} \eta(Y,Z|x) = \eta(\A,\BB)\q \eta(\A,\BB)
\equiv \eta(\A (Y,Z|x) ,\BB (Y,Z|x)  )\,. \ee
Thus, in agreement with   \cite{BHS}, global conformal HS symmetries form the
Lie algebra associated with the Weyl algebra $A_4$ of four pairs of
oscillators. It  contains the $\mathfrak{sp}(8)$ subalgebra of bilinears of
oscillators.
In fact,   $\A^\ua$ and
$\BB_\ub$ can be interpreted as supergenerators
 \be Q^\ua=\A^\ua\q
Q_\ub=\BB_\ub\,, \ee
which, together with
\be T_{\ua\ub} = Q_\ua
Q_\ub\q T^{\ua\ub} = Q^\ua Q^\ub\q T_A{}^\ub = \half\{Q_\ua\,,Q^\ub\}\,, \ee
 form  $\mathfrak{osp}(1,8)$.

The rank-$r$ equations for $r$ species of oscillators $Y^A, Z_A \to Y^A_i,
Z_A^j $ ($i,j=1\ldots r$) can be considered analogously with the Fock space
realization of the rank-$r$ field $|C(Y|x)\rangle$
 \be\label{frankr}
D|C(Y|x)\rangle := (\dr+ W^{(r)}(Y,Z|x))|C(Y|x)\rangle =0\q W^{(r)}(Y,Z|x)=
\sum_{i=1}^r W(Y_i\,,Z^i|x)\,. \ee
 Obviously, these equations are
invariant under the global symmetry with parameters $\eta(\A,\BB)$  valued
in the Weyl algebra $A_{4r}$ generated by
\be\label{AiBi}\A_\ua^i=\A_\ua(Y_i,Z^i|x)    \q\BB_i^\ua=\BB^\ua(Y_i,Z^i|x)
.\quad\ee
 Let
$\mathfrak{hs}(n;\mathbb{C})$ be the complex Lie superalgebra resulting from
$A_n$ via the $\mathbb{Z}_2$ graded commutator $[f\,,g]$ where homogeneous
elements $f(Z,Y)$ are associated with even and odd elements,
$f(Z,Y)=(-1)^{\pi_f} f(-Z,-Y)$. Then the (complexified) rank--$r$ equations
are invariant under $\mathfrak{hs}(4r;\mathbb{C})$. The rank--one HS algebra
$\mathfrak{hs}(4;\mathbb{C})$ belongs to $\mathfrak{hs}(4r;\mathbb{C})$.
Among a number of inequivalent embeddings of $\mathfrak{hs}(4;\mathbb{C})$
into $\mathfrak{hs}(4r;\mathbb{C})$, the principal
embedding where the element $f(Y,Z)\in \mathfrak{hs}(4;\mathbb{C})$ is
represented by $\sum_{i=1}^r f(Y_i,Z_i)$  is most important.

In the $4d$ Minkowski setup it is convenient to use two-component spinor
notation. In these terms, \be\label{4dvar} x^{\ga\pa}=x^{ {n}} \sigma_{n }^{\ga\pb}\,\q Y^A=(y^\ga,\,\,\by^\pb)\q Z_A=(z_\ga,\,\,\bz_\pb)\,, \ee
 where $\sigma_{ {n}}^{\ga\pb}$
   {are    Hermitian} $2\times 2$ {  matrices}
 and $ y^\ga$, $z_\ga$,  $\by^\pb,$ $\bz_\pb$
 are
 conjugated two-component spinor oscillators  with
nonzero commutation relations
\be\label{twocsM}
\dis{[\,z_\gb\,,y^\ga]=\gd^\ga_\gb}\q  [\bz_\pa\,,\by^\pb ]=\gd_\pa^\pb\,.
\ee
 Two-component indices are raised and lowered by the  symplectic
forms $\gvep_{\ga\gb}$ and $\gvep_{\pa\pb}$ \be \label{Cind} A_\gb =A^\ga
\gvep_{\ga\gb}\q A_\pb =A^\pa \gvep_{\pa\pb}\q A^\ga =A_\gb \gvep^{\ga\gb}\q
A^\pa =A_\pb \gvep^{\pa\pb}\,.\qquad
 \ee
 The  algebra $\mathfrak{u}(2,2) $ has connections  $ h^{\ga\pa}\,, \omega_{\ga}{}^{\gb}\,,
\bar{\omega}_{\pa}{}^{\pb}\,$ and  $f_{\ga\pa}$. The Lorentz connection is represented
by the traceless parts $\omega^L_{\ga}{}^{\gb}$ and
$\bar{\omega}^L_{\pa}{}^{\pb}\,$ of  $\omega_{\ga}{}^{\gb}$ and
$\bar{\omega}_{\pa}{}^{\pb}\,,$ respectively,
  while the traces are associated with the gauge fields of
dilatation $b$ and helicity generator $\tilde{b}$
\be \label{btb} b= \half
\Big (\go_\ga{}^\ga + \bar{\go}_\pa{}^\pa\Big )\q \tilde{b}= \half \Big
(\go_\ga{}^\ga -\bar{\go}_\pa{}^\pa\Big )\,. \ee
The $\mathfrak{u}(2,2)$
flatness conditions are
\bee \label{h4} R^{\ga\pb}:&=& \dr h^{\ga\pb}
-\omega_\gga{}^\ga\wedge h^{\gga\pb}- \bar{\omega}_\pga{}^\pb\wedge
h^{\ga\pga} =0\,, \\ \label{b4} R_{\ga\pb}:&=& \dr f_{\ga\pb}
+\omega_\ga{}^\gga\wedge f_{\gga\pb}+ \bar{\omega}_\pb{}^\pga\wedge
f_{\ga\pga} =0\,, \\ \label{o4} R_\ga{}^\gb :&=& \dr\omega_\ga{}^\gb
+\omega_\ga{}^\gga\wedge \omega_\gga{}^\gb -
 f_{\ga\pga}\wedge h^{\pga\gb}=0\,,
\\ \label{oc4} \bar{R}_\pa{}^\pb :&=& \dr\bar{\omega}_\pa{}^\pb
+ \bar{\omega}_\pa{}^\pga\wedge \bar{\omega}_\pga{}^\pb -
f_{\gga\pa}\wedge h^{\gga\pb}=0\,. \eee

Reduction of  unfolded equations (\ref{frank1}) for the massless field ${}
C(y,\by|x)$ to the  $\mathfrak{u}(2,2)
\subset \mathfrak{sp}(8)$ invariant  setup  gives \be \label{Unfsu22} {}\!D_{\mathfrak{u}}^{tw}
{}|C(y,{\bar{y}}|x) \rangle=0\q\ee
 where
\be\label{Dtwsp8} {}\!D_{\mathfrak{u}}^{tw}=
{}\!D_{\mathfrak{su}}^{tw}+ \half\widetilde{b}( y{}^\ga z_\ga -\bar{y}{}^\pa \bz_\pa)
\, \ee
and  the conformal covariant
derivative  ${}\!D_{\mathfrak{su}}^{tw}$ is
\be\label{Dtwc} {}\!D_{\mathfrak{su}}^{tw}= \dr +
 \goL {}_{\ga}{}^{\gb}y{}^\ga z_\gb +
 \bgoL{}_{\pa}{}^{\pb}\bar{y}{}^\pa \bz_ \pb
+ f_{\ga\pa}y^\ga{}\by{}^\pa+
h^{\ga\pa}z_\ga\bz_ \pb
+\half b(  2+  y{}^\ga z_\ga +\bar{y}{}^\pa \bz_\pa)
 \,.\qquad
\ee
The $AdS_4$  description  with the background fields valued in $
\mathfrak{sp} (4)\subset  \mathfrak{su} (2,2)$ results from the Ansatz \be
\label{adsc}
 h^{\ga\pa}=  \lambda e^{\ga\pa} \q
 f_{\ga\pa} =   \lambda  e_{\ga\pa}\q b=\tilde{b}=0\,
\ee and \bee\label{Dtwads} {}\!D_{ads}^{tw}=D^L +\gl
e^{\ga\pa}y_\ga{}\by_\pa+\gl e^{\ga\pa}z_\ga\bz_\pa
 \,\q D^L=  \dr   + \Big ( \goL
{}^{\ga}{}^{\gb}y{}_\ga z_\gb +
 \bgoL{}^{\pa}{}^{\pb}\bar{y}{}_\pa \bz_ \pb
 \Big ).\qquad
\eee
The rank-one unfolded equation in  $AdS_4$  is
\be\label{Adsrank1} {} D_{ads}^{tw}{}|C(y,{\bar{y}}|x) \rangle=0\,.
   \ee

\section{Rank-two equations and covariant oscillators in $AdS_4$}
\label{Rank-two equations}

  In this paper we focus on the     rank--two field $|J(Y|x)\rangle$
  associated with $4d$ conformal conserved currents
built from massless fields of all spins in \cite{GVC}. It satisfies the rank--two
{\emph{current}} equation \bee \label{frank2} D_{2}|J(Y|x)\rangle :=(\dr+
\sum_{i=1,2} W(Y_i\,,Z^i|x))
 |J(Y|x)\rangle=0 \qquad \eee
with $W(Y \,,Z |x)$ (\ref{Wrank1}).

Let us look for a three-form $\go$ closed by virtue of  rank-two
equations
 in the form
 \be \label{defomega}\omega = \langle \Omega|J\rangle\,,
 \ee
 where $\langle \Omega
|$ is a three-form that verifies the equation
\be \label{Omegadualeq}\dr \langle
\Omega | + \langle \Omega |\sum_{i=1,2} W(Y_i\,,Z^i|x)=0\,,
\ee
which,   
together with  (\ref{frank2}),  implies
 \be \dr\omega
=0\,. \ee
On-shell closed forms generate conserved charges
\be Q(\go) = \mathrm{\int}_{\Sigma^3}
\go\,. \ee Conservation means that $Q$ is independent of local variations of $\Sigma^3$
  such as variation of time. Exact $\go$
do not contribute to $Q$ for solutions of the field equations that decrease
fast enough at space infinity. Hence, nontrivial charges $Q$ are associated
with the {\it current cohomology}.

Clearly, a  three-form
 \be \label{clf} \omega(\eta(\A,\BB)) = \langle
\Omega|\eta(\A,\BB)| J\rangle\,,
\ee
where $\eta(\A,\BB)$ is any function of
the oscillators $\A_\ua^i$ and $\BB_i^\ua$    (\ref{AiBi}) with $i=1,2$, is
also closed. Hence, the full space of closed {\it current forms}  is   the space
  of arbitrary functions of  oscillators (\ref{AiBi}).  This freedom
should encode the freedom  in different HS charges. Indeed, as  shown  in
\cite{GSV}, the realization of a rank--two field in terms of bilinears of
rank--one fields gives rise to the full list of conformal gauge invariant
$4d$ conserved currents of all spins identified as generalized Bell-Robinson currents
in \cite{Berends:1985xx}.
However, the freedom in a function of two sets of oscillators $A_\ua^i$ and
$\BB_i^\ua$  is far larger than that in HS symmetries of rank--one equations,
parametrized by a function of the rank--one variables $\A_\ua$ and $\BB^\ua$
(\ref{AaBb}). Hence, we conjectured  in \cite{GVC} that most of the closed
forms (\ref{clf}) are exact,   generating no nontrivial HS charges.

Using  the correspondence between
unfolded and BRST formulations of $Sp(2M)$ invariant HS field equations in the generalized matrix space-time $\M_M$ with any $ M$
the  space of nontrivial parameters $\eta$ and $\widetilde{\eta}$ of bilinear
conserved currents that lead to nonzero
charges $Q_\eta  $ and $\widetilde{Q}_{\widetilde{\eta}}  $
was found in \cite{BRST2010}. The latter  were shown in \cite{OPE} to   generate the
supersymmetric HS algebra  with the doubled number of charges of a given
spin.\footnote{Note however that in \cite{OPE},
where $J$ were realized as bilinears of physical fields, parameters of currents
  satisfied appropriate  symmetry conditions.}  The identification of nontrivial conserved charges in the flat Minkowski
space was also done in \cite{LS} by a different method.

Let us briefly recall the construction of  \cite{BRST2010}.
Consider a generalized space-time $\M_M(X)$ with matrix
  coordinates  $X^{AB}=X^{BA}$,  $A,B=1,\ldots,M$.  $M$-forms
  in $\M_M(X)\times \mathbb{R}^{2M}(U,V)\,$~\cite{gelcur}
\be\label{Warpi} \Phi_\eta( \PPP)\, =\, \half\, \Big(i\hhh\, d\,
X{}^{AB}\f{\p}{\p U^B} +    d\, V{}^A
\Big)^{\,M}\,\eta(\BBB\,,\widetilde{\BBB})\,\PPP(U,\,V\,|X)\Big|_{ \,U=0}\,\qquad
\ee and \be \label{Warpi'} \widetilde{\Phi}_{\widetilde{\eta}}(\PPP)\,= \,\half \,
\Big(i\hhh\, d\, X{}^{AB}\f{\p}{\p V^B} +    d\, U{}^A
\Big)^{\,M}\,\widetilde{\eta}(\BBB\,,\widetilde{\BBB})\,\,\PPP(U,\,V\,|X)\Big|_{V=0}\,\qquad \ee
 are closed by virtue of the current
equation   \cite{tens2} \be \label{rank2eq} \left (\f{\p}{\p
X^{ A B}} - \,i\,\hhh \f{\p^2}{\p U^{( A}\p V^{B)}}\right ) \,
\PPP{}(U,V|X) =0\, \quad \ee
with parameters $\eta$ and $\widetilde{\eta}$ polynomial in
the operators  $\BBB$ and $\widetilde{\BBB} $, respectively,
\bee \label{parametersM}
\BBB_A(U,V|X) &=& \f{\p}{\p U^A} \,\q
 \BBB{}^B(U,V|X) = i\hhh X^{AB}\f{\p}{\p
U^A}+V^B \q \\\label{exparam} \widetilde{\BBB}_A(U,V|X) &=& \f{\p}{\p
V^A}\q  \widetilde{\BBB}{}^B (U,V|X) =\, i\hhh X^{AB}
\f{\p}{\p V^A}+U^B  \,, \qquad \eee which
satisfy\be \label{rank2eqB}  \left[ \f{\p}{\p
X^{ A B}} - \,i\,\hhh \f{\p^2}{\p U^{( A}\p V^{B)}}\, \,,\,\BBB \right ]=\,
\left [ \f{\p}{\p
X^{ A B}} - \,i\,\hhh \f{\p^2}{\p U^{( A}\p V^{B)}}\,\,,\,\widetilde{\BBB} \right] =0\,. \quad \ee
 In  \cite{BRST2010}, it was shown that
 non-exact forms $\Phi $ (\ref{Warpi}) and $\widetilde{\Phi}$ (\ref{Warpi'}) are represented,
respectively, by $\widetilde{\BBB}$-independent ${\eta}(\BBB(U,V|X))$
and  $\BBB$-independent  ${\widetilde{\eta}}(\widetilde{\BBB}(U,V|X))$
which can be interpreted as
 parameters of global HS symmetry transformations generated by the charges
 \bee\label{1}   Q(\PPP_{{\eta(\BBB) }}) =\int \half\, \Big(i\hhh\, d\,
X{}^{AB}\f{\p}{\p U^B}  +    d\, V{}^A
\Big)^{\,M}\,\eta(\BBB\, )\,\PPP(U,\,V\,|X)\Big|_{ \,U=V=0}\q
 \\ \label{2}\widetilde{Q} (\PPP_{
\widetilde{\eta}(\widetilde{\BBB})})=\,\int \half \,
\Big(i\hhh\, d\, X{}^{AB}\f{\p}{\p V^B}   +    d\, U{}^A
\Big)^{\,M}\,\widetilde{\eta}( \widetilde{\BBB})\,\,\PPP(U,\,V\,|X)\Big|_{U=V=0} \,,\qquad \eee
where the integration is over some$M$-dimensional cycle in $\M_M\times \mathbb{R}^{2M}.$
For $J $ bilinear in rank-one fields $C_i$, where $i=1,\ldots, N$ are color indices,
 the  charges generate the HS  algebra \cite{OPE}.

As explained in \cite{gelcur}, the proper restriction of (\ref{1}) and (\ref{2})
at  $M=4$ to $4d$  Minkowski space with local coordinates $x^{\ga\pa}$ gives
Minkowski conserved charges.

In this paper we construct the full list of gauge invariant conserved currents
in $AdS_4$.  Introducing the two-component spinor  oscillators obeying
\be\label{twocsM2}
\dis{[\,z^i_\gb\,,y_j^\ga]=\gd^\ga_\gb}\gd^i_j\q  [\bz^i_\pa\,,\by_j^\pb ]=
\gd_\pa^\pb\gd^i_j\,
\ee
it is convenient to  rescale  $y_2,\by_2\to iy_2,i\by_2$ and $z_2,\bz_2\to
-iz_2,-i\bz_2$ to obtain
from (\ref{frank2})
\bee\label{Dtwads2}
&& \!\!\ls D_{2}{}_{ads}^{tw}|J(Y|x)\rangle=0\q
D_{2}{}_{ads}^{tw}= D_{2}{}^L+\widetilde{W}^{(2)}
 \,\q
 \\ \rule{0pt}{16pt}\label{DLads2} D_{2}{}^L&=&  \dr   + \goL
{}^{\ga}{}^{\gb}(y_1{}_\ga z^1_\gb+y_2{}_\ga z^2_\gb) +
 \bgoL{}^{\pa}{}^{\pb}(\by _1{}_\pa \bz^1_ \pb+\by _2{}_\pa \bz^2_ \pb)
  \q
\\\label{Wrank2}
\rule{0pt}{16pt}\widetilde{W}^{(2)}  &=&
 \gl   e^{\ga\pa}\Big(y_1{}_\ga{}\by_1{}_\pa+  z^1_\ga\bz^1_\pa-y_2{}_\ga{}\by_2{}_\pa-z^2_\ga\bz^2_\pa\Big)\,.\qquad
\eee
Packing the
oscillators $y_i^\ga,$ $\by_i^\pa,$ $z^i_\ga,$ $\bz^i_\pa $
into $ \gdp_\ga^{\nv \nh}$, $  {\bgdp}{}_\pa^{\nv \nh}\, $ with $\nv=-,+$ and
$\nh=\mih,\plh$ by setting
\be\label{gdpbgdp} \beee{llll} \gdp_\ga^{+
\plh}=y_2{}_\ga \q&
\gdp_\ga^{+ \mih}= - y_1{}_\ga \q&
\gdp_\ga^{- \plh}=z^1_\ga\q& 
\gdp_\ga^{- \mih}=z^2_\ga\q\\
\bgdp_\pa^{+ \plh}= -\bz^2_\pa\q& \bgdp_\pa^{+ \mih}=-\bz^1_\pa\q& \bgdp_\pa^{-
\plh}=\by_1{}_\pa\q& \bgdp_\pa^{- \mih}=-\by_2{}_\pa\,, \eeee \ee
one can see that nonzero commutators acquire the form
 \be\label{comreldoup}
[\gdp_\gb^{\nv  \kh} \,,\,\gdp_\ga^{\mv \nh}]=
 \gvep^{\nv \mv  } \gvep^{\kh\nh}\gvep_{\gb\ga}\,\q
 [\bgdp_\pb^{\nv  \kh} \,,\,\bgdp_\pa^{\mv \nh}]=
 \gvep^{\nv \mv  } \gvep^{\kh\nh}\gvep_{\pb\pa}\,,
\ee where indices  are   raised and lowered by $\gvep_{\nv\mv}$ ,
$\gvep_{\nh\mh}$\,, $\gvep^{\nv\mv }$ and $\gvep^{\nh\mh }$ with
\be\label{eps+-} \gvep_{-+}=-\gvep_{+-}=1\q \gvep^{-+}=-\gvep^{+-}=1 \q
  \gvep_{\mih \plh  }= {-}\gvep_{\plh  \mih }=1\q
\gvep^{\mih \plh  }= {-}\gvep^{\plh  \mih }=1 .\ee
From
(\ref{rangle}) it follows
\be\label{oscvac} \langle0|\gdp_\ga^{+ \mh}=0\q
\langle0|\bgdp_\pa^{- \mh}=0\q   \gdp_\ga^{- \mh}| 0\rangle =0 \q
\bgdp_\pa^{+ \mh}| 0\rangle =0\,.\ee 
In these terms, Eq.~(\ref{Wrank2}) takes the form
 \be   \label{W2new}
\widetilde{W} {}^{(2)}(\gdp, \bgdp |x)=
 \gl e^{\ga\pb}\, \gdp_\ga^{\mv\,\nh} \,\bgdp{}_\pb{}_{\mv\,\nh}\,.\quad
\ee

Analogously,
the  covariantly constant oscillators
with respect to the rank--two covariant derivative $
D_{2}{}_{ads}^{tw}$ (\ref{Dtwads2})
 are packed into
 \bee
 \label{packedK}
\!&& \ad_\la^{\nv \nh}( \gdp,\bgdp|x)\q {\bd}{}_\lap^{\nv \nh}( \gdp,\bgdp|x)\,,\qquad \la =1,2;\quad \lap=1,2
\\\nn
\rule{0pt}{20pt}\!&& \left[ D_{2}{}_{ads}^{tw},\ad_\la^{\nv \nh}( \gdp,\bgdp|x)\right]=
\left[ D_{2}{}_{ads}^{tw},\bd_\la^{\nv \nh}( \gdp,\bgdp|x)\right]=0\qquad
 \, \eee
  so
that \be\label{normaAdS} \ad_\la^{m\nh}( \gdp,\bgdp|0) = \gdp_\ga^{m\nh}
\gd_\la^\ga \q \bd_\lbp^{m\nh}( \gdp,\bgdp|0) = \bgdp_\pb^{m\nh}\gd^\pb_\lbp
\,.\qquad \ee
Eq.~(\ref{normaAdS}) guarantees that  $\ad (x)$ and $\bd(x)$
obey analogous  commutation relations
at any $x$
 \be\label{comreladbd}
[\ad _\lb^{\nv  \kh}  (x) \,,\,\ad_\la^{\mv \nh} (x)]=
 \gvep^{\nv \mv  } \gvep^{\kh\nh}\gvep_{\lb\la}\,,\quad
 [\bd_\lbp^{\nv  \kh} (x) \,,\,\bd_\lap^{\mv \nh} (x)]=
 \gvep^{\nv \mv  } \gvep^{\kh\nh}\gvep_{\lbp\lap}\,,\quad
 [\ad_\lb^{\nv  \kh} (x) \,,\,\bd_\lap^{\mv \nh} (x)]=0
  \,.
\ee

This can also be seen in terms of  Killing spinors $\co^\gb(x)$ and
$\si^\pb(x) \, $ of \cite{GVC}, that obey
\be\label{resh} D^L  \co{}^\ga(x)  + \gl e^{\ga\pb}   \si{}_\pb(x)   =0 \,,\quad
 D^L  \si^\pb(x)  +\gl e^{\ga}{}^{\pb}     \co_\ga{}(x) =0\,.\quad
\qquad\ee A basis of the space of solutions of this system is formed by four
independent pairs of spinors $(\co_\la{}^\gb(x),\si_\la{}^\pb(x))$ and
$(\co_\lap{}^\gb(x),\si_\lap{}^\pb(x))$ labeled by  $\la=1,2$ and $ \lap
=1,2$ and obeying  the    conditions
\be\label{reshcon} \co_\la{}^\gb(0)=\gd_\la{}^\gb\q \si_\la{}^\pb(0)=0\q
\co_\lap{}^\gb(0)=0\q \si_\lap{}^\pb(0)=\gd_\lap{}^\pb\,
\ee {}
implying that
$$\overline{\co_\la{}^\gb(x)}=\si_\lap{}^\pb(x)\q
\overline{\si_\la{}^\pb(x)} =\co_\lap{}^\gb(x).$$
A specific  form of the Killing spinors
 depends on a chosen coordinate system.
Covariantly constant oscillators $\ad,\,\,\bd$  (\ref{packedK}) are expressed via the
Killing spinors as \be\label{paramadbd}
 \ad_\la{}^{m\kh}( \gdp,\bgdp|x)=       \co_\la{}^\gga(x)\gdp_\gga{}^{m}{}^{ \kh}
-   \si_\la{}^\pga(x)\bgdp_\pga{}^{m}{}^{ \kh}     \,,\quad
 \bd_\lap{}^{m\kh}( \gdp,\bgdp|x)= -      \co_\lap{}^\gga(x)\gdp_\gga{}^{m}{}^{ \kh}
+   \si_\lap{}^\pga(x)\bgdp_\pga{}^{m}{}^{ \kh} .  \ee

Using (\ref{comreldoup}) and (\ref{W2new})
 we observe that the operators  \bee\label{spv} \vf^{\nv \mv} &=&
\chalf\{\gdp_\gb{}^{\nv}{}_{\mh}\,,\gdp^\gb{}^{\mv}{}^{\mh}\} +\chalf\{
\bgdp_\pb{}^{\nv}{}_{\mh}\,,  \bgdp^\pb{}^{\mv}{}^{\mh}\}\,, \\\label{sph} \hg^{\nh \mh } &=&
\chalf\{\gdp_\gb{}_{\kv}{}^{\nh}\,,\gdp^\gb{}^{\kv}{}^{\mh}\} + \chalf\{
\bgdp_\pb{}_{\kv}{}^{\nh}\,,  \bgdp^\pb{}^{\kv}{}^{\mh}\}\, \eee are
covariantly constant with respect to the rank--two covariant derivative $
D_{2}{}_{ads}^{tw}$ (\ref{Dtwads2})
 forming two mutually commutative $\mathfrak{su}(2)$ algebras. %
 Algebras (\ref{spv}) and (\ref{sph}) will be
referred to as vertical $\slv$  and horizontal $\slh$, respectively.
The standard bases $\vff_j$ of  $\slv$  and $\hgg_j$ of $\slh$ are
\be \label{Hidef} \vff_-=-\vf^{--} =
 z^1{}_\ga  z^2{}^\ga+ \by_1{}^\pa \by^2{}_\pa  \,, \quad
 \vff_+=\vf^{++} =
  y_1{}^\ga y_2{}_\ga+\bz^1{}_\pa \bz^2{}^\pa\,, \quad \vff_0=2\vf^{-+}=2H_2+2H_1
\,,\quad\ee
   \be \nn  \hgg_-=
-\hg^{\mih \mih } =
  y_1{}^\ga\, z^2{}_\ga+\by_2{}^\pa  \bz^1{}_\pa \,, \quad
\hgg_+=\hg^{\plh \plh } =
  y_2{}^\ga\, z^1{}_\ga + \by_1{}^\pa\, \bz^2{}_\pa \,, \quad\hgg_0= 2\hg^{\mih\plh}=2H_2-2H_1\,,\quad
\ee\be  \nn   H_1=\half(y_1{}^\ga z^1{}_\ga -\by_1{}^\pa\bz^1{}_\pa) \q
H_2=\half(y_2{}^\ga z^2{}_\ga -\by_2{}^\pa\bz^2{}_\pa). \qquad\qquad\qquad\ee

The $\slv$  and  $\slh$ act  on $\gdp$ and  $\bgdp$ as follows
\bee\label{doupletrepv} [\vf^{\mv\nv}, \gdp_\gb^{\kv
\mh}]=\half\gvep^{\mv\kv} \gdp_\gb^{\nv  \mh}+ \half\gvep^{\nv\kv}
\gdp_\gb^{\mv  \mh}\q [\vf^{\mv\nv}, \bgdp_\pb^{\kv \mh}]=\half\gvep^{\mv\kv}
\bgdp_\pb^{\nv  \mh}+ \half\gvep^{\nv\kv} \bgdp_\pb^{\mv  \mh} \q\\\nn
[\hg^{\mh\nh}, \gdp_\gb^{   \mv \kh}] =\half\gvep^{\mh\kh} \gdp_\gb^{\mv
\nh}+\half\gvep^{\nh\kh} \gdp_\gb^{\mv \mh}\q [\hg^{\mh\nh}, \bgdp_\pb^{
\mv \kh}] =\half\gvep^{\mh\kh} \bgdp_\pb^{\mv \nh}+\half\gvep^{\nh\kh}
\bgdp_\pb^{\mv  \mh} \,.\qquad \eee
From here it follows that  $\slh$ and $\slv$ act  on the
 oscillators $\ad\,$ and $\,\bd$  (\ref{paramadbd}) analogously
\bee\label{doupletrepkil} 
[\vf^{\mv\nv}, \ad_\gb^{\kv\mh}]
=\half\gvep^{\mv\kv} \ad_\gb^{\nv  \mh}
+ \half\gvep^{\nv\kv}\ad_\gb^{\mv  \mh}\q [\vf^{\mv\nv}, \bd_\pb^{\kv \mh}]
=\half\gvep^{\mv\kv}\bd_\pb^{\nv  \mh}
+ \half\gvep^{\nv\kv}  \bd_\pb^{\mv  \mh} \q
\\\nn
[\hg^{\mh\nh}, \ad_\gb^{   \mv \kh}] =\half\gvep^{\mh\kh} \ad_\gb^{\mv\nh}
+\half\gvep^{\nh\kh} \ad_\gb^{\mv  \mh}\q [\hg^{\mh\nh}, \bd_\pb^{   \mv
\kh}] =\half\gvep^{\mh\kh} \bd_\pb^{\mv \nh}+\half\gvep^{\nh\kh} \bd_\pb^{\mv\mh} \,.\qquad
\eee
Indeed,  being covariantly constant,  $\vf\in\slv$ and $\hg\in\slh$
keep the same form in terms of
  $\ad_\la^{\mv \mh}$ and $ \bd_\lap^{   \mv \kh}$,
 \be\label{spvab} \vf^{\nv \mv} =
\chalf\{\ad_\lb{}^{\nv}{}_{\mh}\,,\ad^\lb{}^{\mv}{}^{\mh}\} +\chalf\{
\bd_\lap{}^{\nv}{}_{\mh}\,,  \bd^\lap{}^{\mv}{}^{\mh}\}\,, \ee
\be\label{sphab} \hg^{\nh \mh } =
\chalf\{\ad_\lb{}_{\kv}{}^{\nh}\,,\ad^\lb{}^{\kv}{}^{\mh}\} + \chalf\{
\bd_\lap{}_{\kv}{}^{\nh}\,,  \bd^\lap{}^{\kv}{}^{\mh}\}\,. \ee

\section{On-shell de Rham
cohomology of $AdS_4$  currents} \label{Currents AdS}
 Let us show that the following
three-forms $\langle \Omega^{\mh}{}^{\kh}|=\langle \Omega^{\kh}{}^{\mh}|=
 \langle 0 | \Omega^{\mh}{}^{\kh}$ with
\be\label{OMEGA} 
    \Omega^{\mh}{}^{\kh}      : =
 \Hh^{\ga\pa}  \{\gdp_\ga^{- \mh} \, \bgdp_\pa{}_{-}{}^{ \kh}+
 \gdp_\ga^{- \kh} \, \bgdp_\pa{}_{-}{}^{ \mh}\}       \, \q
 {\Hh}^{\ga \delta^\prime}=-\f{1}{3}  e^{\ga}{}_\pa   e^\gb{}^\pa e_{\gb}{}^{\delta^\prime}\,
  \ee
 obey (\ref{Omegadualeq}). Indeed, using   (\ref{oscvac}) and
\be {\Hh}^{\eta \pb}\,
e^{\gga}{}^{\pgga}=\gvep^{\eta\gga}\gvep^{\pb\pgga}{\mathbf{H}}\q
{\mathbf{H}}:=\chalf e_{\eta\sigma^\prime} {\Hh}^{\eta \sigma^\prime}
 \,, \ee 
Eqs.~(\ref{comreldoup}) and (\ref{W2new}) yield
\bee \label{Omegadualproof} \dr  \Big\langle
\Omega^{\mh}{}^{\kh} \big|
 +   \Big\langle\Omega^{\mh}{}^{\kh} \big|
\widetilde{W }{}^{(2)}(\gdp, \bgdp |x)=
 \Big\langle 0 \big| \big[\widetilde{W }{}^{(2)}(\gdp, \bgdp |x)\,,\,
 \Omega^{\mh}{}^{\kh} \big]
 +   \Big\langle\Omega^{\mh}{}^{\kh} \big|
\widetilde{W }{}^{(2)}(\gdp, \bgdp |x)\,&&
\\ \nn=\Big\langle0 \big| \widetilde{W }{}^{(2)}(\gdp, \bgdp
|x) \,\Omega^{\mh}{}^{\kh}   =
 \gl{\mathbf{H}}\Big\langle 0\big |
 \gdp_\gb^{-\,\nh} \,\bgdp{}_\pb{}_{-\,\nh}   \,
\{\gdp^\gb{}^{- \mh} \, \bgdp^\pb{}{}_{-}{}^{ \kh}+
 \gdp^\gb{}^{- \kh} \, \bgdp^\pb{}{}_{-}{}^{ \mh}\}
\,.&&
 \eee
The last term  is zero since being symmetric in $\kh$ and $\mh$
it turns out to be antisymmetric by virtue of (\ref{comreldoup}). Then
 the form
\be\label{closekil} \go =\big\langle  \Omega^{\mh}{}^{\kh}\,\big|
 \eta_{\mh}{}_{\kh}(\ad,\bd)
 \big| \PPP{}(Y|x)\big\rangle
  \ee
 is closed for any
 $\eta_{\mh}{}_{\kh}$ provided
that $\PPP{}(Y|x)$ verifies  current equation  (\ref{Dtwads2}).

 The central fact of the analysis of the current cohomology is that   each of the forms
  \be\label{exact!kil}
\go_\la^{\mv}{}^{\mh}=\big\langle  \Omega^{\mh}{}^{\kh}\,\big|
\ad_\la{}^{\mv}{}_{\kh} \, \eta(\ad,\bd)
\big|  \PPP{}(Y|x)\big\rangle
 \q\go_\lbp^{\mv}{}^{\mh}=\big\langle  \Omega^{\mh}{}^{\kh}\,\big|
 \bd_\lbp{}^{\mv}{}_{\kh} \,\eta(\ad ,\bd ) \big| \PPP{}(Y|x)\big\rangle\qquad
  \ee
 is exact provided that $\big|\PPP(Y|x)\big\rangle$ solves  (\ref{Dtwads2}). 

For instance, let us prove that  $\go_\la{}^{-\,\mh}$ (\ref{exact!kil}) with $\eta=1$
is exact (other cases are  analogous).
Indeed, using that
 \be\nn \qquad\qquad{H}^{\ga
\gb}\, e^{\gga}{}^{\delta^\prime}= \gvep^{\ga\gga}{\Hh}^{\gb \delta^\prime}
+\gvep^{\gb\gga}{\Hh}^{\ga \delta^\prime} \q H^{\ga\gb} =
e^{\ga}{}_\pa \, e^\gb{}^\pa\,, \ee
 we obtain by virtue of  Eq.~(\ref{resh})  
\bee\label{detaIhat} && \ls \ls\ls\dr
E_b^{\!-\!}{}^{\mh}:=\dr\big\langle 0\big|  H^{\ga\gb}  {\co}{}_\lb {}_\ga \gdp_\gb^{\!-\!}{}^{
\mh}\big|J(Y|x)\big\rangle\!=\!
 \left\langle 0\big |  H^{\ga\gb} (\dr{\co}{}_\lb {}_\ga )\gdp_\gb^{\!-\!}{}^{ \mh} \!-\!
   H^{\ga\gb}  {\co}{}_\lb {}_\ga \gdp_\gb^{\!-\!}{}^{ \mh}\widetilde{W} {}^{(2)}(\gdp, \bgdp |x)
   \big|J(Y|x)\right\rangle\\
&&\ls  \!=\!
    \gl\left\langle 0\big |   {3}   \Hh^{\gb}{}^{\pb}
        {\si}_\lb {}_\pb(x)  \gdp_\gb^{\!-\!}{}^{ \mh}
    \!+\!  \big(\gvep^{\gm\ga} \Hh^{\gb\pn}\!+\!\gvep^{\gm\gb}
     \Hh^{\ga\pn}\big)
     \co{}_\lb {}_\ga    \gdp_\gb{}^{\!-\!}{}^{ \mh}
   \big( \gdp_\gm{}^{\!+\!}{}^{ \nh}
 \bgdp{}_\pn{}_{\!+\! \nh}\!+\!
 \gdp_\gm{}^{\!-\! \nh}  \bgdp_\pn{}_{\!-\! \nh}\big) \big|J(Y|x)\right\rangle\,.
\nn
\eee
Neglecting the term with $\bgdp{}_\pn{}_{\!+\! \nh}$ by virtue of
(\ref{oscvac}) and using (\ref{paramadbd}) along with
 \be\label{skobkiOM}
    [\Omega^{\mh}{}^{\kh} , \ad_\la{}^{-}{}_{\kh}]  =[
 \Hh^{\ga\pa}  \{\gdp_\ga^{- \mh} \, \bgdp_\pa{}_{-}{}^{ \kh}+
 \gdp_\ga^{- \kh} \, \bgdp_\pa{}_{-}{}^{ \mh}\} , \ad_\la{}^{-}{}_{\kh}]  =
   -{3}  \Hh^{\ga\pa}  {\si}_\la{}_\pa(x)     \gdp_\ga^{- \mh}\,,   \ee
(\ref{detaIhat}) yields
\bee
&&\ls \ls \dr E_b^{\!-\!}{}^{\mh}\!=\! \gl\left\langle 0\big |    -[\Omega^{\mh}{}^{\kh} ,
\ad_\lb{}^{\!-\!}{}_{\kh}]
   \!+\!
      \Hh^{\gb\pn}\!
     \co{}_\lb {}^\gm    \gdp_\gb{}^{\!-\!}{}^{ \mh}
     \gdp_\gm{}^{\!-\! \nh}  \bgdp_\pn{}_{\!-\! \nh}
      +\!      \Hh^{\gb\pn}\co{}_\lb {}_\gb    \gdp^\gm{}^{\!-\!}{}^{ \mh}
     \gdp_\gm{}^{\!-\! \nh}  \bgdp_\pn{}_{\!-\! \nh}  \big|J(Y|x)\right\rangle
  \\ \nn
&& \ls \!=\! \gl\left\langle 0\big |    - \Omega^{\mh}{}^{\kh}
\ad_\lb{}^{\!-\!}{}_{\kh}
\!+\!
\Hh^{\ga\pa} \co_\la{}^\gga(x)\gdp_\gga{}^{-}{}_{ \kh}
 \{\gdp_\ga^{- \mh} \, \bgdp_\pa{}_{-}{}^{ \kh}+
 \gdp_\ga^{- \kh} \, \bgdp_\pa{}_{-}{}^{ \mh}\}
\right.  \\ \nn
&&  \left.  \!+\!    \Hh^{\gb\pn} {\co}{}_\lb {}^\gm \gdp_\gm^{\!-\! \nh} \{
  \gdp_\gb^{\!-\!}{}^{ \mh} \bgdp_\pn{}_{\!-\! \nh}
\!+\!\gdp_\gb^{\!-\!}{}_{ \nh} \bgdp_\pn{}_{\!-\!  }{}^{ \mh}\} \big|J(Y|x)\right\rangle
  \!=\!   -  \gl \big\langle 0\big| \Omega^{\mh}{}^{\kh}   \ad_\lb{}^{\!-\!}{}_{\kh}\big |J(Y|x)\big\rangle
   \,,\qquad  \quad \eee
where we used that $\gdp_\ga^{\!-\! \nh}
\gdp_\gb^{\!-\!}{}_{ \nh}= \half \gvep_{\ga\gb}\gdp_\gamma^{\!-\! \nh}
\gdp{}^\gamma{}^{\!-\!}{}_{ \nh} $.

Analogously one can show that other forms  (\ref{exact!kil})  are exact.
   An important consequence of this fact and   commutation relations
   (\ref{comreladbd}) is that the following forms are also exact:
 \be\label{exactrel}
\big\langle \Omega^{\mh\nh}\big|\{\ad_\la{}^{m}{}_{\kh},\ad_\lb{}^{k \kh}\}
   \eta(\ad,\bd) \big|\PPP{}\rangle
     ,\quad
    \langle  \Omega^{\mh}{}^{\nh}\big|\{\bd_\lap{}^{m}{}_{\kh},\bd_\lbp{}^{k \kh}\}
   \eta(\ad,\bd)\big| \PPP{}\rangle
     ,\quad
     \langle  \Omega^{\mh}{}^{\nh}\big|\{\ad_\la{}^{m}{}_{\kh},\bd_\lbp{}^{k \kh}\}
   \eta(\ad,\bd)\big| \PPP{}\rangle\,.
     \quad
 \ee

For instance, since  $ \langle \Omega^{\mh}{}_{\nh}\big|(\ad_\la{}^{m\,\nh}\,
   \eta(\ad,\bd) \big|\PPP{}\rangle $ is exact, the form
 \bee\label{exactrelpr} && \langle \Omega^{\mh}{}_{\nh}\big|
(\ad_\la{}^{m\,\nh}\, \,\ad_\lb{}^{k\,\kh}- \ad_\lb{}^{k\,\nh}\,
\,\ad_\la{}^{m\,\kh})
   \eta(\ad,\bd) \big|\PPP{}\rangle=
   \qquad\\ \nn&&\half
   \langle \Omega^{\mh}{}_{\nh}\big|(\{\ad_\la{}^{m\,\nh}\, ,\,\ad_\lb{}^{k\,\kh}\}-
\{\ad_\lb{}^{k\,\nh}\,,\,\ad_\la{}^{m\,\kh}\})   \eta(\ad,\bd)\big|
\PPP{}\rangle =\half\gvep^{\nh \kh} \big\langle\Omega^{\mh}{}_{\nh}\big|
\{\ad_\la{}^{m}{}_{\ph}\, ,\,\ad_\lb{}^{k\,\ph}\}
\big|\PPP\big\rangle\,\qquad \eee
is exact. The proof for the other forms in
(\ref{exactrel}) is  analogous.

This fact admits the following interpretation. Various
bilinears in $\ad_\la{}^{m}{}^{\nh}\,$ and $ \,\bd_\lbp{}^{k\,\kh} $
form a Lie algebra $\mathfrak{sp}({16})$ while
 \be\label{o8}\G{}_{\la\,\lb}^{m\,k}=\half\{\ad_\la{}^{m}{}_{\kh},\ad_\lb{}^{k\,\kh}\}\q
\G{}_{\la\,\lbp}^{m\,k}=\half\{\ad_\la{}^{m}{}_{\kh},\bd_\lbp{}^{k\,\kh}\}\q
 \G{}_{\lap\,\lbp}^{m\,k}=\half\{\bd_\lap{}^{m}{}_{\kh},\bd_\lbp{}^{k\,\kh}\}
\ee
   form a Lie algebra   $\mathfrak{o}(4,4)$ 
   that commutes with the  horizontal $\slh$ (\ref{sph})
 acting  on the  hatted  indices.
For  parameters $\eta$ polynomial in oscillators,
 factorization of  generators (\ref{o8}) allows us to factor out any
combination of oscillators containing antisymmetrization of a pair of  the hatted
Latin indices.\footnote{Beyond the space of polynomials the situation is
different because, formally, it follows that all nontrivial
$\mathfrak{o}(4,4)$-modules should be factored out. However, this `exact'
representation for conserved currents
turns out to be space-time nonlocal, containing infinitely
many derivatives, giving rise to quasi-exact representations analogous to
those considered  in \cite{Prokushkin:1999ke}.}
The remaining forms $\widetilde{\omega}$ (\ref{closekil}) have
 totally symmetrized  hatted indices.

 To describe  such forms consider the space $\mathbf{P}_{AdS}$ of
 {\it preforms}
    \be\label{Lambda} \Omega^{ \nh,\mh} \eta(\ad_\la^{m \kh} , \bd_\lap^{n \ph}) \ee
  with
totally symmetrized  hatted indices, including those of $\Omega^{ \nh,\mh}$.
 Clearly,  $\slh$ (\ref{sph}) leaves  $\mathbf{P}_{AdS}$
 invariant.  Since any $\slh$ highest vector in $\mathbf{P}_{AdS}$ that is symmetric in  hatted indices has the form
 \be\label{omegatil} \Omega^{ \plh   \plh  } \eta (\ad_\la^{m \plh  } , \bd_\lap^{n \plh  }), \ee
   $\mathbf{P}_{AdS}$ is a span of vectors
   \be\label{P+}
 \tgo =  \mathrm{ad}^N_{  \hgg_-} \left(\Omega^{\plh \plh}
    \eta (\ad_\la^{m \plh} ,  \bd_\lap^{n \plh} )\right)\q \mathrm{ad}_x(y)=[x,y]\,\q\hgg_-
    \in \slh
     \ee
with various $N\ge0$, \ie
\be\label{Padsum}
\mathbf{P}_{AdS}=\sum_{N}
\oplus \tgo  \,.
     \ee

Now we observe
 that   the  Cartan   element   $\hgg_0\in\slh$
(\ref{Hidef})   annihilates  both $|0\rangle$
and $\langle 0|$. Hence,
 \be \label{centranih}
\langle 0\,\big| [\hgg_0\,, \tgo    \PPP]\big|\,0\rangle=0  \ee for any $\tgo  $
(\ref{P+})   and $\PPP$. This implies that  $\langle 0\,\big|  \tgo    \PPP \big|\,0\rangle $
 can be nonzero
only if
   \be \label{centranihK}
[\hgg_0\,,\tgo  \PPP] =0 . \ee
Evidently  any current $\PPP(Y|x)$ and  parameter
$\eta   (\ad ,  \bd )$ can be decomposed as
\be \label{Jsvsh}\PPP=\sum_{\gmm}\PPP_\gmm\q\eta=\sum_{\gnn}\eta_\gnn\q
[\hgg_0\,,\PPP_{\gmm } ] =\gmm\PPP_{\gmm } \q
 [\hgg_0\,,\eta_{\gnn } ] =\gnn\eta_{\gnn }\q \mu,\nu\in {\mathbb Z} \,.
 \ee
 Since \be [\hgg_0,\ad_\la^{m \plh}]=\ad_\la^{m \plh}\q [\hgg_0,\bd_\lap^{m \plh}]=\bd_\lap^{m \plh}
 \q [\hgg_0,\Omega^{\plh \plh}]=2\Omega^{\plh \plh}\q
 \ee
then
\be\left(\mathrm{ad}_{  \hgg_-}\right)^N \left(\Omega^{\plh \plh}
    \eta_\gnn  (\ad_\la^{m \plh} ,  \bd_\lap^{n \plh} )\right)=0\quad
    \mbox{if}\quad N>2+\gnn \,.
    \ee
Hence for  given $  \PPP_{\gmm }$ and
  $\eta_{\gnn }$
  the condition $
\big\langle   0\big|\gtgo_N (\eta_{\gnn }  )
  \PPP_{\gmm }{}(Y|x)\big|0\rangle \ne 0
$ demands
  \be\label{helic} -2-\gnn\le \gmm\le2+\gnn\q
2 + \gnn+  \gmm =2N\q  \ee
determining  $N$  unambiguously in terms of $\nu$ and $\mu$.

Preforms $ \gtgo_{gen}\in \mathbf{P}_{AdS}$ can be represented as
\be\label{gensum}
 \gtgo_{gen}=\sum_{N}
  \left(\mathrm{ad}_{  \hgg_-}\right)^N \left(\Omega^{\plh \plh}
    \eta   (\ad_\la^{m \plh} ,  \bd_\lap^{n \plh}  )\right)c_N (\hgg_0) \,
     \ee
with some $\hgg_0$-dependent coefficients $c_N (\hgg_0)$ encoding
the freedom in normalization of charges.

The simplest choices $c_N=\f{(\pm 1)^N}{N!} \alpha{}^\pm(\hgg_0)$ with some
$\alpha{}^\pm ( \hgg_0)$ yield the generating functions
\be \label{genP+}\gtgo^\pm _{gen} =
  \exp \pm \left(\mathrm{ad}_{  \hgg_-}\right) \left(\Omega^{\plh \plh}
    \eta^+  (\ad_\la^{m \plh} ,  \bd_\lap^{n \plh}  )\right) \alpha{}^\pm(\hgg_0)
      \ee
appropriate for the description of currents carrying positive and negative
helicities.

Indeed, as mentioned in \cite{OPE}, any current $J(Y|x)$  can also be  decomposed into a sum of currents
$J^h(Y|x)$ of different {\it current helicities} $h$, that satisfy \be
[\Hh, J^h(Y|x)] = h J^h(Y|x)\,, \ee where  the {\it current helicity
operator}  is expressed in terms of $\hgg_j$ (\ref{Hidef})  as
\be
\label{helicitytoka}
 \Hh=  \chalf\left({\hgg_++\hgg_-}\, \right).
 \ee


Using   Eqs.~(\ref{Hidef}), (\ref{doupletrepv}), (\ref{doupletrepkil}),
(\ref{OMEGA}) and  the Taylor expansion $ f(x+y)=\exp(x\f{\p}{\p y})f(y) $,
Eq.~(\ref{genP+}) yields
    \bee\label{Omega+}
  \gtgo ^\pm_{gen}=  \widetilde{\Omega}^\pm
      \eta^\pm( \widetilde{\ad}_\la^{m}{}_{ \pm}   ,  \bd_\lap^{n}{}_{ \pm}  |\hgg_0)
 ,\eee
 where  $\eta^\pm( a  ,  b  |\hgg_0)=\eta^\pm  (a ,b) \, \alpha{}^\pm(\hgg_0)\,,$
  \be\widetilde{\ad}_\la^{m}{}_{ \pm}=\ad_\la^{m \plh} \pm\ad_\la^{m \mih}\q
  \widetilde{\bd}_\lap^{n}{}_{ \pm}=\bd_\lap^{n \plh}\pm \bd_\lap^{n \mih}\q
   \widetilde{\Omega}^\pm     = \Omega^{\plh\plh}+\Omega^{\mih\mih} \pm
   \Big(\Omega^{\mih\plh}+\Omega^{\plh\mih}\big)
                .\quad
  \ee

The preforms $\gtgo ^\pm_{gen}$ (\ref{Omega+}) generate the space of
 nontrivial closed forms
 \bee\label{resultconscur+}
 \go_+ =\big\langle   \widetilde{\Omega}^+\,\big|
      \eta_+( \widetilde{\ad}_\la^{m}{}_{ +}   ,  \widetilde{\bd}_\lap^{n}{}_{ +} |\hgg_0 )
  \big| \PPP{}(Y|x)\big\rangle\q
    \\\label{resultconscur-}
 \go_- =\big\langle   \widetilde{\Omega}^-\,\big|
      \eta_-( \widetilde{\ad}_\la^{m}{}_{ -}   ,
      \widetilde{\bd}_\lap^{n}{}_{ -}|\hgg_0)
  \big| \PPP{}(Y|x)\big\rangle\qquad\,\,\,
  \eee
(and hence conserved currents) equivalent to the space
of  parameters depending on a single set
of oscillators   $\widetilde{\ad}_\la^{m}{}_{ +}   ,  \widetilde{\bd}_\lap^{n}{}_{ +}$ in
(\ref{resultconscur+}) and
$\widetilde{\ad}_\la^{m}{}_{ -}   ,  \widetilde{\bd}_\lap^{n}{}_{-}$ in
(\ref{resultconscur-})\!.
 As mentioned above,
the dependence on $\hgg_0$ in (\ref{resultconscur+}), (\ref{resultconscur-})
encodes the freedom in normalization of the original bilinear current~(\ref{bilinear}).

Since
\be\label{curhelcom}
[\Hh,\widetilde{\ad}_\la^{m}{}_{ \pm} ]=\pm\half\widetilde{\ad}_\la^{m}{}_{ \pm} \q[\Hh,
\widetilde{\bd}_\lap^{n}{}_{ \pm}]=\pm\half\widetilde{\bd}_\lap^{n}{}_{ \pm}\q
 [\Hh,\widetilde{\Omega}^\pm ]=\pm\widetilde{\Omega}^\pm ,\qquad
\ee
$ \go_+$ and $ \go_-$ (\ref{resultconscur+}) depend on the parameters
carrying non-negative and non-positive current helicities, respectively.

\newcommand{\rhor}{\varrho}
\newcommand{\brhor}{\bar{\varrho}}
\newcommand{\epsr}{\epsilon}

 Reformulating the result  in the notations
 analogous (up to some rescalings) to those of \cite{GVC} with
 \be\label{reskl}z^j{}_\ga\to \p^j{}_\ga=\f{\p}{\p y_j^\ga}\q \bz^j{}_\pa\to \p^j{}_\ga=\f{\p}{\p \by_j^\pa}\q
 {\p_\pm{}_\ga}\sim\p^1{}_\ga\pm
\p^2{}_\ga \q{y_\pm{}_\ga}\sim y_1{}_\ga\pm y_2{}_\ga ,\quad \etc \q\ee
\be\nn\left\{
\widetilde{\ad}_\la^{m}{}_{+} ,
\widetilde{\bd}_\lap^{n}{}_{+}\right\} \to\left\{  {\rhor},\bar{\rhor}\right\}\q
\left\{
\widetilde{\ad}_\la^{m}{}_{-} ,
\widetilde{\bd}_\lap^{n}{}_{-}\right\} \to\left\{  {{\epsr}},\bar{\epsr}\right\}\,\q\ee
nontrivial charges   are represented by the closed three-forms
\bee\label{concur2helres}\Omega_\eta(\PPP)&=& \Hh^{\ga\pa} {\p_ -{}_\ga} {\p _-{}_\pa}
\eta(\rhor,\brhor|\hgg_0) \PPP{}(y^\pm,\by^\pm|x)\Big|_{{y^\pm=\by^\pm=0}}
,
\\\nn\Omega_{\widetilde{\eta}}(\PPP)&=&
\Hh^{\ga\pa}{\p_+{}_\ga} {\p _+{}_\pa} \widetilde\eta (\epsr,\bar{\epsr}|\hgg_0)
\PPP{}(y^\pm,\by^\pm|x)\Big|_{{y^\pm=\by^\pm=0}} , \eee
where
 \bee\nn
   \rhor_- {}_ \la
    =    \co_\la{}^\ga(x)\p_-{}_\ga
+   \si_\la{}^\pa(x)  \by^+{}_\pa \q\,
   {\rhor}^+{}_ \la  =      \co_\la {}^\ga(x) y^+_\ga
+  \si_\la{}^\pa(x)  \bp_-{}_\pa        \q\, \, \\ \nn
 \!\!{\epsr}{\,}^- {}_ \la  =     \co_\la {}^\ga(x) y^-_\ga +  \si_\la {}^\pa(x)
\bp_+{}_\pa      \q\,\,\,\,
{\epsr}_+{}_ \la
=   \co_\la{}^\ga(x)\p_+{}_\ga +   \si_\la{}^\pa(x)  \by^-{}_\pa       \qquad\,\,
     \eee
  and  $\co^\gb(x)$ and
$\si^\pb(x) \, $ are Killing spinors
(\ref{resh}), (\ref{reshcon}). $\bar{\rhor}$ and $\bar{\epsr}$ are complex conjugated to $ {\rhor}$ and $ {\epsr}$, respectively.
In these variables
 the current helicity operator $\Hh$ (\ref{helicitytoka})
 is \be\label{helicitytoka'}
 \Hh=\half\Big(\by_+^\pa \bp^+_\pa+ y_+^\ga \p^+_\ga-\by_-^\pa \bp^-_\pa- y_-^\ga \p^-_\ga\Big)
. \ee

 For bilinear currents
   \be\label{bilinear}
   \PPP(y^\pm\,\by^\pm|x)= 
   C^+_{p_+ }(y^++y^-,\by^++\by^-| x )C^-_{p_-}(y^+-y^-,\by^+-\by^- | x)\,  \q 
   \ee
where  the fields $C_{p_\pm}^\pm(y\,,\by|x)$  carry helicities $p_\pm $   and  solve  rank--one equations
(\ref{Adsrank1})
$$ {}D_{ads}^{tw} C^+_{p_+}(y ,\by | x )=0\q D_{ads}^{tw}C^-_{p_-}(iy ,i\by | x )=0\,,$$
 Eqs.~(\ref{concur2helres})  yield  the following
  two closed forms announced in \cite{GVC}:
\be\label{concur2helCC}
 \Hh^{\ga\pa}\f{\p}{\p y^-{}^\ga}\f{\p}{\p \by^-{}^\pa} \eta (\rhor,\bar\rhor
| p_+-p_-) C^+_{p_+}(y^++y^-,\by^++\by^-| x )C^-_{p_-}(y^+-y^-,\by^+-\by^- |
x)\Big|_{{y^\pm=\by^\pm=0}}
 ,\quad
\ee\be\nn
 \Hh^{\ga\pa}\f{\p}{\p y^+{}^\ga}\f{\p}{\p \by^+{}^\pa} \widetilde{\eta}
(\epsr,\bar\epsr|p_+-p_-  )
 C^+_{p_+}(y^++y^-,\by^++\by^-| x )C^-_{p_-}(y^+-y^-,\by^+-\by^- | x)
 \,\Big|_{{y^\pm=\by^\pm=0}}\,,\quad
 \ee
which represent two generating functions for gauge invariant conformal HS current cohomology in $AdS_4$.
By virtue of (\ref{curhelcom}) and (\ref{reskl}) the charges \bee\nn
 {Q^+}_\eta = \int  \Hh^{\ga\pa} {\p_ -{}_\ga} {\p _-{}_\pa}
\eta(\rhor,\brhor, p_+-p_-)  C^+_{p_+}(y^++y^-,\by^++\by^-| x )
C^-_{p_-}(y^+-y^-,\by^+-\by^- | x)\,\Big|_{{y^\pm=\by^\pm=0}}
,\quad
\\\nn
{Q^-}_{\widetilde{\eta}}=\int \Hh^{\ga\pa}{\p_+{}_\ga} {\p _+{}_\pa}
\widetilde{\eta} (\epsr,\bar{\epsr}, p_+-p_-)
 C^+_{p_+}(y^++y^-,\by^++\by^-| x )C^-_{p_-}(y^+-y^-,\by^+-\by^- | x)\,
 \Big|_{{y^\pm=\by^\pm=0}}\,\,\,\quad\eee are
supported by the parameters of non-negative and non-positive
current helicities, respectively.
This list of charges   matches the
 space of parameters of the $4d$ conformal HS symmetry algebra
 as discussed in \cite{GVC}.

 \section{Minkowski  current cohomology }
\label{Currents Min}

Analogous results for flat Minkowski space
 announced in \cite{GVC} can be obtained as follows.
In the limit $\lambda\to0$, appropriately rescaled  Minkowski  current
equations,
 resulting from
 (\ref{Dtwads2}) along with (\ref{Wrank2}) and (\ref{DLads2}),  take the form
 (see, e.g.,  \cite{{GVC}} and references therein)
\bee\label{DtwM}   D_{2}{}_{Mnk}^{tw}|J(Y|x)\rangle&=&0\q
 D_{2}{}_{Mnk}^{tw}= \dr +\widetilde{W}{}_{Mnk}^{(2)} \q\\\label{Wrank2M}
\widetilde{W}{}_{Mnk}^{(2)}  &=&
   e^{\ga\pb}\, \gdp_\ga^{-\,\nh} \,\bgdp{}_\pb{}_{-}{}_{\,\nh}
 \,.\eee
One can see that $\vf^{+-}\in \slv$ (\ref{spv}) and the full algebra $\slh$
(\ref{sph}) commute with $D_{2}{}_{Mnk}^{tw}$. Minkowski Killing spinors in the
Cartesian
coordinate system with $D^L=\dr$
\renewcommand{\co}{\tilde{c}}
\renewcommand{\si}{\tilde{s}}
\be\label{minkil} \co_\la^\ga=\gd_\la^\ga\q \si_\lap^\pa=\gd_\lap^\pa\q
\co_\lap^\ga=-x^\ga{}^\pa \gvep_{\pa \lap}\q \si_\la^\pa=-x^\ga{}^\pa
\gvep_{\ga \la}\,\qquad \ee obey
\be\label{reshM} \dr \co{}_\la^\ga      =0 \q
 \dr  \si_\la^\pb   + e^{\ga}{}^{\pb}     \co_\la{}_\ga{}  =0\,\q
 \dr  \si{}_\lap^\pga       =0 \q
 \dr  \co_\lap^\ga   + e^{\ga}{}^{\pb}     \si_\la{}_\pb{}  =0\,.
\qquad \ee
 Hence, one can choose the following  basis of 
 covariantly constant oscillators
 \bee\label{paramadbdM}
  \adm_\gga{}^{+\,\kh}= \,\,\,      \co_\la{}^\gga(x)\gdp_\gga{}^{+}{}^{ \kh}
-   \si_\la{}^\pga(x)\bgdp_\pga{}^{+}{}^{ \kh}     \,\q\rule{10pt}{0pt}&&
\adm_\gga{}^{-\,\kh}=       \co_\la{}^\gga(x)\gdp_\gga{}^{-}{}^{ \kh}
\,,
\\
 \bdm_\lap{}^{-\,\kh}= -      \co_\lap{}^\gga(x)\gdp_\gga{}^{-}{}^{ \kh}
+   \si_\lap{}^\pga(x)\bgdp_\pga{}^{-}{}^{ \kh}
\q&& \bdm_\lap{}^{+\,\kh}= 
\si_\lap{}^\pga(x)\bgdp_\pga{}^{+}{}^{ \kh}\, .  \eee
Analogously to the $AdS_4$ case, in terms of (\ref{reskl}), this gives that  Minkowski
nontrivial charges   are fully represented by the following closed
three-forms
\renewcommand{\rhor}{\gx}
\renewcommand{\brhor}{\bar{\gx}}
\renewcommand{\epsr}{\chi}
\bee\label{concur2hel} \Hh^{\ga\pa} {\p_ -{}_\ga} {\p _-{}_\pa}
\eta(\rhor,\brhor|\hgg_0) \PPP{}(y^\pm,\by^\pm|x)\Big|_{{y^\pm=\by^\pm=0}}
,
\\\nn
\Hh^{\ga\pa}{\p_+{}_\ga} {\p _+{}_\pa} \eta (\epsr,\bar{\epsr}|\hgg_0)
\PPP{}(y^\pm,\by^\pm|x)\Big|_{{y^\pm=\by^\pm=0}} , \eee
 where $J$ satisfies Minkowski   current equations (\ref{DtwM}) and
   \be \label{parflop+}
  {\chi}{}{}_+{}_\ga=\frac{\p }{\p  y^+{}^\ga}\,,\,\,
  \bar{\chi}{}{}{}_+{}_\pb=\frac{\p }{\p  \by^+{}^\pb}
 \,,\,\,
 {\chi}{}{}{}^-{}^\ga= y^-{}^\ga \,- \,x{}^{\ga\pb}  \frac{\p }{\p  \by^+{}^\pb} \, ,\,\,\
\bar{\chi}{}{}{}^-{}^\pa= \by^-{}^\pa \,- \,x{}^{\gb\pa}  \frac{\p }{\p
y^+{}^\gb} \,,
 \ee
\be \label{parflop}
  {\xi}{}{}_-{}_\ga=\frac{\p }{\p  y^-{}^\ga}\,,\,\,
  \bar{\xi}{}{}{}_-{}_\pb=\frac{\p }{\p  \by^-{}^\pb}
 \,,\,\,
 {\xi}{}{}{}^+{}^\ga= y^+{}^\ga \,-  \,x{}^{\ga\pb}  \frac{\p }{\p  \by^-{}^\pb} \, ,\,\,\
\bar{\xi}{}{}{}^+{}^\pa= \by^+{}^\pa \,-  \,x{}^{\gb\pa}  \frac{\p }{\p
y^-{}^\gb} \,. \ee
As announced in \cite{GVC},  (\ref{concur2hel})
represents two generating functions for the nontrivial current
cohomology in Minkowski space, which are supported by the parameters
of non-negative and non-positive current helicities, respectively.

 \section*{Acknowledgments}
This research was supported in part by the RFBR Grant No. 14-02-01172.
We are grateful to Dmitry Sorokin for  useful comments.

  \addtocounter{section}{1}
\addcontentsline{toc}{section}{\,\,\,\,\,\,\,References}

\end{document}